\documentclass[aps,twocolumn,showpacs,preprintnumbers,amsmath,amssymb]{revtex4-1}
\usepackage{graphicx,psfrag}
\usepackage{color}
\definecolor{navyblue}{rgb}{0.3,0.3,1}
\definecolor{purple}{rgb}{0.6,0,0.5}
\usepackage[colorlinks=true, pdfstartview=FitV, linkcolor=blue, 
citecolor=blue,urlcolor=navyblue]{hyperref}

\begin{document}

\title{Stellar properties and nuclear matter constraints} 

\author{Mariana Dutra$^1$, Odilon Louren\c{c}o$^2$ and D\'ebora P. Menezes$^3$} 

\affiliation{$^1$Departamento de Ci\^encias da Natureza - IHS, Universidade Federal
Fluminense, 28895-532 Rio das Ostras, RJ, Brazil \\
$^2$Departamento de Ci\^encias da Natureza, Matem\'atica e Educa\c c\~ao, CCA, 
Universidade Federal de S\~ao Carlos, 13600-970 Araras, SP, Brazil \\
$^3$Depto de F\'{\i}sica - CFM - Universidade Federal de Santa Catarina, Florian\'opolis 
- SC - CP. 476 - CEP 88.040 - 900 - Brazil}
\date{\today}

\begin{abstract}
We have analyzed stellar properties of the relativistic mean-field (RMF) parametrizations 
shown to be consistent with the recently studied constraints related to nuclear matter, 
pure neutron matter, symmetry energy and its derivatives [Dutra {\it et al}., Phys. Rev. C 
90, 055203 (2014)]. Our results show that only two RMF parametrizations do not allow the 
emergence of the direct Urca process, important aspect regarding the evolution of a 
neutron star. Moreover, among all approved RMF models, fourteen of them produce neutron 
stars with maximum masses inside the range \mbox{$1.93\leqslant M/M_\odot\leqslant 2.05$}, 
with $M_\odot$ being the solar mass. Only three models yield maximum masses above this 
range and a discussion on the inclusion of hyperons is presented. Finally, we have 
verified that the models satisfying the neutron star maximum mass constraint do not 
observe the squared sound velocity bound, namely, $v_s^2 < 1/3$, corroborating recent 
findings. However, the recently proposed \mbox{$\sigma$-cut} scheme can make the RMF 
models consistent with both constraints depending on the isoscalar-vector interaction of 
each parametrization.
\end{abstract}

\pacs{21.30.Fe, 21.65.Cd, 26.60.Kp, 24.10.Jv}

\maketitle

\section{Introduction}

Historically, around 1930, the first nuclear physics model known as the liquid drop 
model~\cite{liquiddrop} and the semi-empirical mass formula presented by Bethe and 
Weizs\"acker~\cite{semiempirical}, established the grounds for the study of nuclei 
properties and nuclear structure. These two models are so close to each other in basic 
ideas, that their nomenclature is very often  mixed up. Both are parameter dependent 
models and their underlying fitting procedure was later used in the development of many 
other models.  Nowadays, around~500 hundred nonrelativistic (Skyrme-type) and 
relativistic models are available in the literature and largely used in describing 
different features, from nuclear to hadron spectroscopy, from heavy-ion collisions to 
neutron star properties. The vast majority of the models are parameter dependent based on 
the fitting of nuclear bulk matter. These phenomenological models generally rely on the 
calculation of equations of state~(EoS), which relate pressure, energy density and 
temperature at a given particle number density subject to nuclear forces. 

In the last years, detailed analyses of nonrelativistic and relativistic models were 
performed~\cite{PRC85-035201,PRC90-055203}. From such studies, it was suggested that 
the proliferation of models and the production of new parameter sets with a limited range 
of application should not be encouraged. To be more specific, 
in~\cite{PRC85-035201}, 240 different Skyrme model parameterizations were confronted 
with experimentally and empirically derived constraints and only 16 were shown to satisfy 
all of them. Other~263 relativistic mean-field (RMF) parametrizations were also 
analysed in~\cite{PRC90-055203} and once again, only a small number of them, 35 to 
be exact, were shown to satisfy adequately chosen constraints. It is important to say 
that three different sets of constraints were used in~\cite{PRC90-055203}, all of them 
related to symmetric nuclear matter, pure neutron matter, symmetry energy and its 
derivatives. They differed one from the other in the choice of validity ranges of certain 
quantities and in the level of restriction. In a recent study~\cite{jeremy}, the authors 
also constrained the EoS of Skyrme and RMF models to those predicted from chiral effective 
field theory including two- and three-nucleon forces effects.

In~\cite{PRC90-055203}, the relativistic models were divided into~7 families, namely, 
linear finite range models (Walecka-type models~\cite{walecka}, type~1), non-linear 
$\sigma$ models (Boguta-Bodmer models~\cite{boguta}, type~2), non-linear $\sigma$ and 
$\omega$ models with a self-quartic interaction in the $\omega$ field (type~3),
non-linear $\sigma$ and $\omega$ terms and cross terms involving these fields (type~4), 
density dependent models~\cite{NPA656-331} with couplings adjusted to nuclear properties 
(type~5), non-linear point coupling models~\cite{PRC46-1757} (type~6) and models with 
$\delta$ mesons (type~7). Thirty of the approved models are of type~4, two are of 
type~5, one of type~6, and two of type~7 being both density dependent. In the 
present work, we go one step further and confront all the approved models with 
observational astrophysical properties. Until not very long ago (2010), all 
mean-field models adapted to stellar conditions (charge neutrality and chemical 
equilibrium) that resulted in maximum stellar masses larger than $1.44M_\odot$ were 
acceptable. After two massive stars were discovered, namely, the PSR~J1614-2230 with 
mass of \mbox{$(1.97\pm0.04)M_\odot$}~\cite{nature467-2010}, and the PSR~J0348+0432 with 
mass of \mbox{$(2.01\pm0.04)M_\odot$}~\cite{science340-2013}, many parameter dependent 
models were {\it re-tuned} so that they could describe maximum masses in these 
ranges. Other mechanisms capable of stiffening the EoS and hence increasing the maximum 
stellar mass were also proposed as in~\cite{Weiss,luiz2014,voskre2015}, but we will 
discuss them after we present our results. 

The aim of the present work is to check whether the 35 RMF parametrizations that 
were shown to satisfy the nuclear matter constraints in~\cite{PRC90-055203} also satisfy 
the criterion of producing massive stars. The same EoS are also used to investigate  the
direct Urca process and to calculate the sound velocity.

\section{Formalism and Results}

To obtain neutron star macroscopic properties, the necessary steps are
the following ones: 1) the EoS for hadronic matter is joined with the
EoS for free leptons; 2) the conditions of charge neutrality and
chemical equilibrium are enforced; 3) the BPS equation of state \cite{BPS}
for low densities is added to the EoS for hadrons and leptons; 4) the
resulting EoS is used as input to the Tolman-Oppenheimer-Volkoff
equations \cite{TOV}, which are the differential equations for the 
structure of a static, spherically symmetric star in hydrostatic
equilibrium. All these steps are well known and the details are given
in many papers and books, as for instance, in \cite{Glen, Haensel}.
Next, only electrons and muons are considered in the leptonic EoS
since we just consider the zero temperature deleptonized phase of the
stellar evolution.

As a remark, we remind the reader that clusterized and possibly the pasta phase 
matters can also be present in the inner crust of the star. However, in the present calculation, 
it is important to point out that after we join the BPS EoS, used to
describe the outer crust, to the hadronic one obtained 
from the RMF models, the TOV equations are solved. In doing so, an interpolation is used 
and the exact values in between these two EoSs loose their strict meaning. In our 
calculations, the lowest baryonic density obtained from the relativistic models is 
approximately $\rho=0.05$~fm$^{-3}$ and the BPS points start respectively at 
$\rho=0.008907$~fm$^{-3}$ (baryonic density), $\varepsilon=0.04253$~fm$^{-4}$ (energy 
density) and $p=0.00006987$~fm$^{-4}$ (pressure) and go up to a very low baryonic 
density, namely $\rho=0.1581\times10^{-10}$~fm$^{-3}$, and corresponding energy density 
and pressure. Any different matter structure existing in between these
two densities is washed out when the TOV equations are used. Hence, the detailed 
structure of the inner crust is not relevant for the present analysis. In 
our analysis, we start with nucleons only in the hadronic sector and the results are 
presented next.

\subsection{Nucleonic matter}

\begin{figure}[!htb]
\centering
\psfrag{M/Msun}[b][b][1.0]{$M/M_\odot$}
{\includegraphics[scale=0.48]{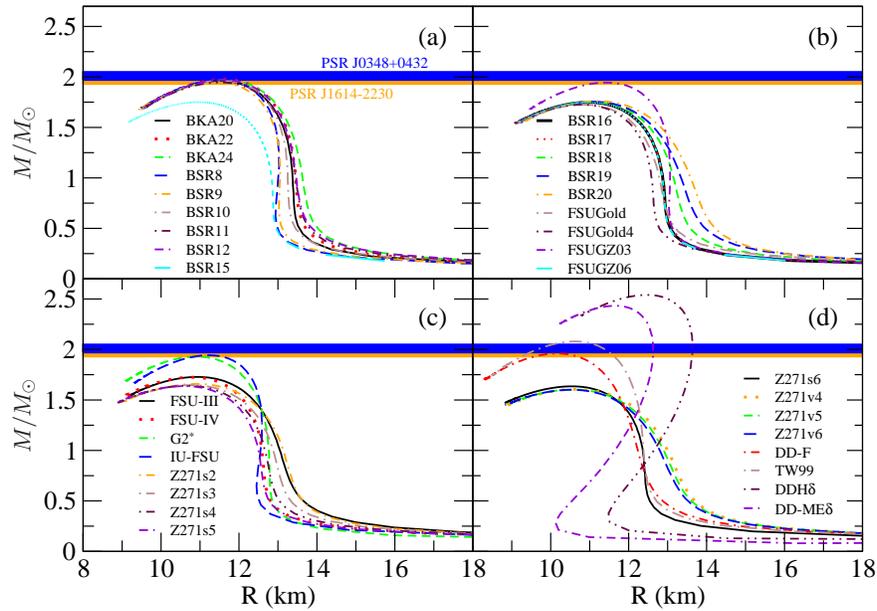}}
\caption{(Color online) Neutron star mass-radius relation. Horizontal bands indicate the 
masses of PSR~J1614-2230~\cite{nature467-2010} (orange) and 
PSR~J038+0432~\cite{science340-2013} (blue).}
\label{fig1}
\end{figure}

In Table~\ref{tab1} we display some of the neutron star properties obtained from the 
solution of the TOV equations after the 35 models approved in~\cite{PRC90-055203} are 
used as input. The first 30 results correspond to models of type 4, described in the 
Introduction and the respective Lagrangian densities can be obtained from the references 
listed in the second column. The last 4 results come from density dependent models and in 
the last 2, the scalar isovector $\delta$ meson is also included. Regarding the only 
point-coupling model approved according to Ref.~\cite{PRC90-055203}, the FA3 one, we 
remark that due to its very particular behavior at high density regime, namely, a fall in 
the curve $p\times\varepsilon$ at around $\varepsilon=4.1$~fm$^{-4}$ ($p$ is the pressure 
and $\varepsilon$ is the energy density), it was not possible to generate a mass radius 
curve indicating a maximum mass and, consequently, no other quantities presented in 
Table~\ref{tab1}. Therefore, we have discarded such a model from our analysis. In 
Fig.~\ref{fig1}, the corresponding mass radius curves are plotted, alongside with the 
bands that represent astrophysical constraints of the order of $2M_\odot$ on the 
stellar masses. From these results, one can immediately see that the models BKA20, BKA22, 
BKA24, BSR8, BSR9, BSR10, BSR11, BSR12, FSUGZ03, G2*, IU-FSU, and DD-F are inside the 
maximum mass constraint of~\cite{nature467-2010,science340-2013}, and that only 3 models 
yield maximum masses above the bands shown in Fig.~\ref{fig1}, namely, TW99, DDH$\delta$ 
and DD-ME$\delta$. It is a very well known fact that hyperons are expected to exist in 
the core of neutron stars and that when they are included in the EoS, it becomes softer 
and the corresponding maximum stellar mass decreases considerably. Hence, just these 3 
models will be selected for the study of hyperonic matter. Before we proceed to perform 
this analysis, we look at three other important aspects related to neutron stars: their 
radii, the direct Urca process (DU) and the sound velocity.

As far as neutron star radii are considered, there is a bit of controversy in 
their acceptable values, as discussed in the literature. In Ref.~\cite{schwenk}, for 
example, the authors constrained the radii of the canonical $1.44M_\odot$ neutron star to 
the range of $9.7\leqslant R_{1.44M_\odot}\leqslant 13.9$~km, based on a chiral effective 
theory analysis. In our study, we observe that all values are within such a range (see the 
fifth column of Table~\ref{tab1}). In Ref.~\cite{steiner1}, on the other hand, a limit of 
$12$~km is found for $R_{1.44M_\odot}$, while in Ref.~\cite{steiner2} the limit is 
$13.1$~km. Another calculation, discussed in Ref.~\cite{steiner3}, based on a Bayesian analysis, 
results in radii of all neutron stars in the range of $10.9$~km and $12.7$~km, while even 
significantly smaller radii are predicted in Refs.~\cite{guillot1,guillot2}, where the 
ranges found are of $9.1^{1.3}_{-1.5}$~km and $9.4\pm1.2$~km, respectively, obtained from 
analyses of five quiescent low-mass x-ray binaries. Finally, it is also possible to infer 
neutron star radii from the knowledge of the symmetry energy slope at the saturation 
density ($\rho_o$) due to the correlation between such quantities, as suggested in 
Refs.~\cite{rafael,corr}. Some of these ranges and other ones extracted from other recent 
publications are summarized in Fig.~10 of Ref.~\cite{fortin}. As one can see, definite 
values for the neutron star radii are not yet established, and the results obtained from 
the RMF parametrizations analyzed in the present work (most of them in the range of 
$12\leqslant R_{1.44M_\odot}\leqslant 13$~km) are compatible with several of the 
aforementioned predictions.
\onecolumngrid
\textcolor{white}{asdfas}
\begin{table}[!htb]
\scriptsize
\caption{Neutron star main properties: maximum stellar masses in terms of $M_\odot$ 
($M_{\rm max}/M_\odot$), the radius of the corresponding star ($R$), the radius of a star 
with a $1.44M_\odot$ mass ($R_{1.44M_\odot}$) and the central energy density of the 
maximum mass star ($\varepsilon_c$). Properties related to the direct Urca process: mass 
of the star in terms of $M_\odot$ at the onset of the DU process ($M_{\rm DU}/M_\odot$), 
related baryonic density ($\rho_{\rm DU}$) and proton fraction ($Y_{\rm DU}$).}
\centering
\begin{tabular}{l|c|c|c|c|c|c|c|c}
\hline\hline
Model & Ref. & $M_{\rm max}/M_\odot$ & $R$~(km) & $R_{1.44M_\odot}$~(km) & 
$\varepsilon_c$(fm$^ {-4}$) & $M_{\rm DU}/M_\odot$ & $\rho_{\rm DU}$ (fm$^{-3}$) &  
~~~~$Y_{\rm DU}$~~~~ \\
\hline
BKA20 & \cite{PRC81-034323} & $1.960$  & $11.522$ & $13.191$ & $6.177$ & $1.065$ & 
$0.311$ 
& $0.133$  \\
\hline
BKA22 & \cite{PRC81-034323} & $1.975$  & $11.601$ & $13.262$ & $6.095$  & $1.000$ & 
$0.294$ & $0.133$ \\ 
\hline
BKA24 & \cite{PRC81-034323} & $1.968$  & $11.608$ &  $13.367$ & $6.160$ & $0.911$ & 
$0.271$ & $0.132$  \\
\hline
BSR8 & \cite{PRC76-045801}  & $1.969$  & $11.503$ &  $12.970$ & $6.090$ & $1.410$ & 
$0.405$ & $0.135$  \\
\hline
BSR9 & \cite{PRC76-045801}  & $1.944$  & $11.419$ &  $12.958$ & $6.240$ & $1.313$ & 
$0.385$ & $0.135$  \\
\hline
BSR10 & \cite{PRC76-045801} & $1.963$  & $11.533$ &  $13.108$ & $6.137$ & $1.149$ & 
$0.335$ & $0.134$  \\
\hline
BSR11 & \cite{PRC76-045801} & $1.946$  & $11.504$ & $13.208$ & $6.264$ & $0.980$ & 
$0.294$ 
& $0.133$  \\
\hline
BSR12 & \cite{PRC76-045801} & $1.970$  & $11.580$ &  $13.252$ & $6.160$ & $1.011$ & 
$0.300$ & $0.133$  \\
\hline
BSR15 & \cite{PRC76-045801} & $1.750$  & $10.969$ & $12.483$ & $6.844$ & $1.250$ & 
$0.420$ 
& $0.136$  \\
\hline
BSR16 & \cite{PRC76-045801} & $1.748$  & $10.968$ & $12.494$ & $6.893$ & $1.231$ & 
$0.414$ 
& $0.136$  \\
\hline 
BSR17 & \cite{PRC76-045801} & $1.750$  & $10.989$ & $12.494$ & $6.902$ & $1.119$ & 
$0.372$ 
& $0.135$  \\
\hline
BSR18 & \cite{PRC76-045801} & $1.751$  & $11.040$ &  $12.662$ & $6.873$ & $1.028$  & 
$0.338$ & $0.134$ \\
\hline
BSR19 & \cite{PRC76-045801}  & $1.754$  & $11.102$ & $12.773$ & $6.828$ & $0.923$ & 
$0.303$ & $0.133$ \\
\hline
BSR20 & \cite{PRC76-045801}  & $1.760$  & $11.194$ & $12.972$ & $6.772$ & $0.841$ & 
$0.267$ & $0.132$  \\
\hline
FSU-III & \cite{PRC85-024302}  &  $1.728$  & $10.934$ & $12.502$ & $7.011$ & $1.058$ & 
$0.362$ & $0.135$  \\
\hline
FSU-IV & \cite{PRC85-024302}   &  $1.725$  & $10.797$ & $12.220$ & $7.061$ & $1.477$  & 
$0.572$ & $0.138$ \\
\hline
FSUGold & \cite{PRL95-122501}  &  $1.725$  & $10.842$ & $12.337$ & $7.090$ & $1.304$ & 
$0.467$ & $0.136$ \\
\hline
FSUGold4 & \cite{NPA778-10} &  $1.725$  & $10.788$ & $12.224$ & $7.110$ & $1.479$ & 
$0.572$ & $0.138$ \\
\hline
FSUGZ03 & \cite{PRC74-034323}  &  $1.944$  & $11.418$ & $12.963$ & $6.240$ & $1.314$ & 
$0.385$ & $0.135$ \\
\hline
FSUGZ06 & \cite{PRC74-034323}  &  $1.748$  & $10.961$ & $12.494$ & $6.893$ & $1.244$ & 
$0.419$ & $0.136$ \\
\hline
G2$^*$ & \cite{PRC74-045806}   &  $1.929$  & $10.907$ & $12.551$ & $6.959$ & $1.197$ & 
$0.390$ & $0.135$ \\
\hline
IU-FSU & \cite{PRC82-055803}   &  $1.943$  & $11.228$ & $12.563$  & $6.348$ & $1.776$ & 
$0.614$ & $0.138$ \\
\hline
Z271s2 & \cite{PRC66-055803}   &  $1.658$  & $10.910$ & $12.464$ & $7.098$ & $1.079$ & 
$0.366$ & $0.135$ \\
\hline
Z271s3 & \cite{PRC66-055803}   &  $1.647$  & $10.751$ & $12.238$ & $7.301$ & $1.301$ & 
$0.488$ & $0.137$ \\
\hline
Z271s4 & \cite{PRC66-055803}  & $1.640$ & $10.669$ & $12.070$ & $7.347$ & $1.470$ & 
$0.637$ & $0.138$ \\
\hline
Z271s5 & \cite{PRC66-055803}  & $1.637$ & $10.590$ & $11.952$ & $7.424$ & $1.562$ & 
$0.789$ & $0.140$ \\ 
\hline
Z271s6 & \cite{PRC66-055803}  & $1.635$ & $10.534$ & $11.859$ & $7.469$ & $1.607$ & 
$0.934$ & $0.140$ \\
\hline
Z271v4 & \cite{PRC66-055803}  & $1.606$ & $10.639$ & $12.085$ & $7.590$ & $1.272$ & 
$0.500$ & $0.137$ \\
\hline
Z271v5 & \cite{PRC66-055803}  & $1.603$ & $10.572$ & $11.984$ & $7.667$ & $1.501$ & 
$0.748$ & $0.140$ \\
\hline
Z271v6 & \cite{PRC66-055803}  & $1.601$ & $10.514$ & $11.902$ & $7.743$ & $1.585$ & 
$1.028$ & $0.141$ \\
\hline
DD-F & \cite{PRC74-035802}  & $1.960$ & $10.173$ & $11.880$ & $7.981$ & $-$ & $-$ & $-$ \\
\hline
TW99 & \cite{NPA656-331}  & $2.080$ & $10.613$ & $12.254$ & $7.239$ & $-$ & $-$ & $-$ \\
\hline
DDH$\delta$ & \cite{NPA732-24} & $2.540$ & $12.460$ & $13.282$ & $4.917$ & $2.000$ & 
$0.569$ & $0.138$ \\
\hline
DD-ME$\delta$ & \cite{PRC84-054309} & $2.433$ & $11.657$ & $12.296$ & $5.520$ & $2.112$ & 
$0.766$ & $0.140$ \\
\hline \hline
\end{tabular}
\label{tab1}
\end{table}
\twocolumngrid

\onecolumngrid
\textcolor{white}{asdfas}
\twocolumngrid

The direct Urca (DU) process, $n\to p+e^-+ \bar{\nu}_e$~\cite{urca} sheds light on an 
important aspect of the evolution of neutron stars. The cooling of the star by neutrino 
emission can occur relatively fast when the proton fraction exceeds a critical value 
$x_{\rm \scriptscriptstyle DU}$~\cite{urca}, evaluated in terms of the leptonic fraction 
as~\cite{PRC74-035802} $x_{\rm \scriptscriptstyle DU} = \frac{1}{1 + (1 + x_e^{1/3})^3}$, 
where 
$x_e = \rho_e/(\rho_e + \rho_\mu)$ is the electron leptonic fraction, with $\rho_e$ 
and $\rho_{\mu}$ being the densities of electrons and muons, respectively. Cooling 
rates of neutron stars~\cite{veronica} seem to indicate that this fast cooling process 
does not occur, except in stars with a mass larger than $1.5M_\odot$~\cite{PRC74-035802}. 
Hence, from the results in Table~\ref{tab1}, one can see that all selected models could 
give rise to the DU process. In the same table, we show the results for the star mass 
obtained when the proton fraction crosses the line of the electron fraction.  We also 
display the results of the corresponding baryonic density and the proton fraction. While 
most of the analyzed models cross the line that establishes the onset of the DU process 
when the proton fraction is of the order of $0.14$, the densities can vary substantially. 
Most models cross this line for very low densities, but there are some exceptions, 
when the crossing takes place at around half the saturation density or even at higher 
densities (Z271s5, Z271s6, Z271v5, Z271v6, DD-ME$\delta$). In~\cite{rafael}, it is shown 
that the DU process is related to the density dependence of the symmetry energy and hence, 
to the isovector channel of the EoS and to its slope: the larger the slope of the 
symmetry energy, which corresponds to a harder symmetry energy, the smaller the onset 
density because larger proton fractions are favored in the system. Hence, our analyses 
of~32 of the models show that the fast cooling can indeed take place if the stars 
are described by them. On the other hand, the parametrizations DD-F and TW99 do not 
support this possibility because the proton fraction never crosses the electron leptonic 
fraction.
\onecolumngrid
\textcolor{white}{asdfas}
\begin{figure}[!htb]
\centering
{\includegraphics[scale=0.48]{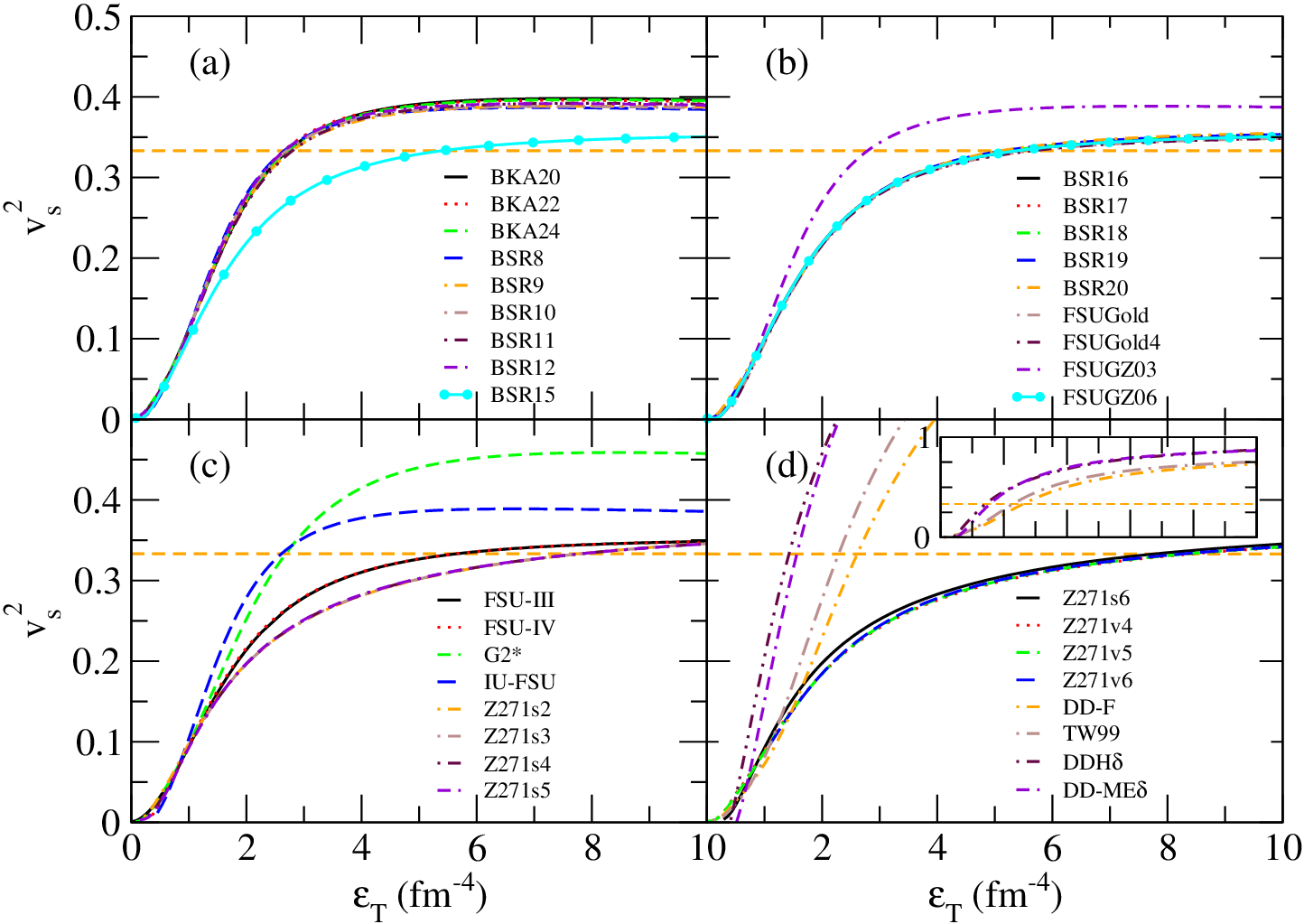}}
\caption{(Color online) Squared sound velocity for nuclear matter in beta 
equilibrium. Dashed line represents the limit $v^2_s = \frac{1}{3}$.}
\label{fig2}
\end{figure}
\twocolumngrid

\subsection{Sound velocity constraint}

We now move our attention to the calculation of the sound velocity, which can be easily 
obtained from the EoS, since for matter in beta equilibrium it is given by 
$v_s^2=dp_T/d\varepsilon_T$, where $p_T$ and $\varepsilon_T$ are the total pressure 
and total energy density of the system, respectively. In Fig.~\ref{fig2} we show the 
behaviour of the squared sound velocity as a function of the total energy density. 
Causality constrains $v_s$ to the light velocity in vacuum, $c$, which we take as $1$ in 
the present work. Although only four models (DD-F, TW99, DDH$\delta$ 
and DD-ME$\delta$) reach quite high values, they never exceed $0.9$, as one can 
see in the inset of Fig.~\ref{fig2}{\color{blue}(d)}, and even though, only at very high 
energy densities, around $\varepsilon_T=10$~fm$^{-4}$, which are higher than the 
central energy densities in stars described by these models, as seen in Table~\ref{tab1}. 

We still remark here that results derived from QCD~\cite{soundQCD} and also 
according to Ref.~\cite{paulo}, another limitation to the sound velocity, given by 
$v_s^2=1/3$, is confirmed by several classes of strongly coupled theories with 
gravity duals. However, if one examines, for instance, Fig.~5 of 
Ref.~\cite{luiz2014}, where a study on the effects of meson-hyperon coupling constants on 
the onset of hyperons in dense nuclear matter is performed, one can see that at the onset 
of every individual hyperon, the sound velocity shows a peak, decreases and starts 
increasing again. If instead of the appearance of new hyperons, the EoS suffered a 
transition to another phase, which could be a mixed phase of hyperons and quarks or a 
phase containing only quarks (see Ref.~\cite{euro}, as an example), the decrease in the 
sound velocity could be even more abrupt. The conclusions drawn in Ref.~\cite{paulo} refer 
to matter in one phase only. Had the authors considered a phase transition (not discarded 
by observational constraints), the squared sound velocity could reach values higher than 
$1/3$, decrease considerably at the appearance of the new phase, and then increase up to 
values around $1/3$, related to the limit imposed by QCD.

Our results show that most of the models reach values larger than $1/3$ at 
energy densities below the stellar central energy density. The exceptions are the class of 
models Z271, where the squared sound velocity reaches $1/3$ at energy densities 
slightly higher than the corresponding star central energy density. We do not have enough 
statistics to produce a histogram as the one shown in~\cite{paulo}, but our findings 
corroborates the statement in~\cite{paulo}, namely, that are unlikely models with 
acceptable behavior at low densities capable of producing maximum masses around 
$2M_\odot$, and, simultaneously, satisfying the bound $v_s^2 < 1/3$, since here 
we are not taking into account the possibility of the system undergoing phase 
transitions. Notice that in the case of the models Z271, compatible with sound velocity 
bound, the maximum masses are all lower than $2M_\odot$, as we can see in 
Table~\ref{tab1}.

Based on these results, let us examine those parametrizations in which the maximum 
neutron star mass is below the minimum value of $1.93M_\odot$ established in 
Ref.~\cite{nature467-2010}, namely, \mbox{BSR15-BSR20}, \mbox{FSU-III}, \mbox{FSU-IV}, 
\mbox{FSUGold}, \mbox{FSUGold4}, \mbox{FSUGZ06}, \mbox{Z271s2-Z271s6}, and
\mbox{Z271v4-Z271v6}. A valid attempt to make such parametrizations consistent with 
the the constraint of \mbox{$1.93\leqslant M/M_\odot\leqslant 2.05$} for the maximum 
neutron star mass, is to modify the models at high density limit ($\rho>\rho_o$) in such 
way that their saturation properties are not altered, i. e., quantities such as binding 
energy, saturation density, effective mass, incompressibility, symmetry energy, etc., are 
kept the same for each parametrization. This procedure ensures that results presented by 
such models under nuclear matter constraints at $\rho=\rho_0$, investigated in 
Ref.~\cite{PRC90-055203}, remain valid. 

In Ref.~\cite{voskre2015}, the authors proposed the called \mbox{$\sigma$-cut} scheme in 
which they add in the $U(\sigma)$ potential of the RMF models, the function $\Delta 
U(\sigma)=\alpha{\rm ln}\{1 + \exp[\beta(f-f_{\mbox{\tiny s.core}})]\}$, where 
$f=g_\sigma\sigma/M_N$ and $f_{\mbox{\tiny s.core}}=f_0+c_\sigma(1-f_0)$. The value of $f$ 
at the saturation density is $f_0$; $\alpha$, $\beta$ and $c_\sigma$ are constants; 
$g_\sigma$ regulates the strength of the attractive interaction, denoted by the scalar 
field $\sigma$; and $M_N$ is the nucleon rest mass. In this scheme, it is possible to 
avoid the decreasing in the density dependence of the effective nucleon mass, 
$M^*=M_N-g_\sigma\sigma$, through the function $\Delta U(\sigma)$. Such a decreasing is 
present in the original RMF models where $\Delta U(\sigma)=0$. It is responsible by 
softening the EoS (since the attraction is increasing), avoiding the system to reach 
higher values for the maximum neutron star mass. According to Ref.~\cite{voskre2015}, the 
necessary condition that needs to be satisfied in order to prevent $M^*$ of decreasing in 
the \mbox{$\sigma$-cut} scheme, is expressed in terms of the constants $\alpha$ and 
$\beta$ by $\alpha\beta^2\gg 6\times10^{-3}M_N^4\rho/\rho_o$. Since this condition is 
verified, the constant $c_\sigma$ controls at which density, denoted by $\rho_*$, the 
decreasing in $M^*$ stops. In order to ensure that the bulk parameters of the models are 
preserved, it is natural to choose $c_\sigma$ in such way that $\rho_*>\rho_0$.

\begin{figure}[!htb]
\centering
{\includegraphics[scale=0.33]{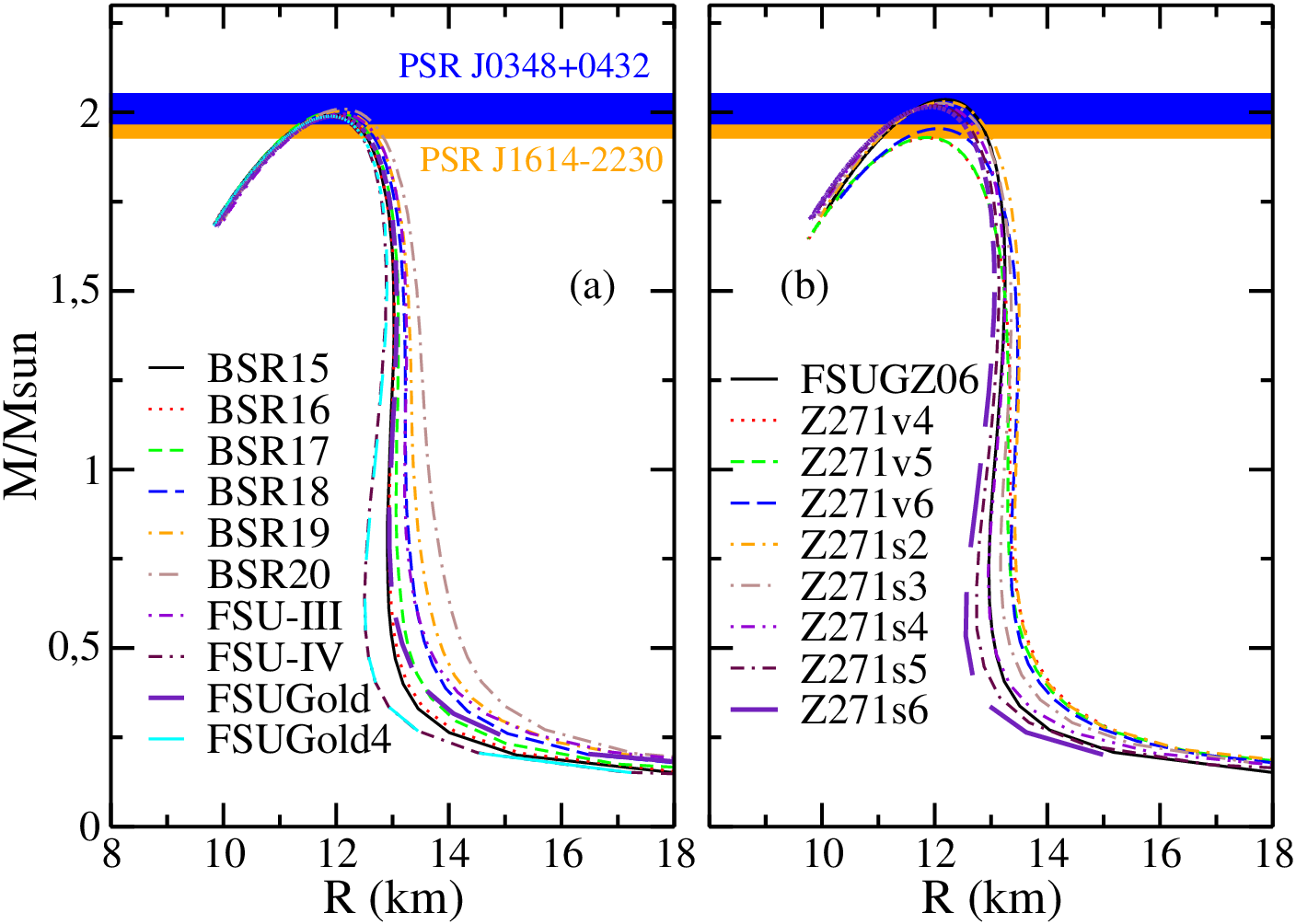}}
\caption{(Color online) $M\times R$ for models presenting \mbox{$M<1.93M_\odot$} in 
Table~\ref{tab1}. Here we used (see text) $c_\sigma=0.35$ and $0.31$, respectively, for 
the BSR and FSU parametrizations. For Z271v4, Z271v5 and Z271v6 ones, we chose 
$c_\sigma=0.145$, $0.14$ and $0.12$, respectively.}
\label{soft}
\end{figure}

Let us consider here the values of $\alpha=4.822\times10^{-4}M_N^4$ and 
$\beta=120$ used in Ref.~\cite{voskre2015}, and apply the \mbox{$\sigma$-cut} scheme in 
the models presenting $M<1.93M_\odot$. By properly choosing the $c_\sigma$ 
values, it is possible to generate the neutron star mass-radius relation depicted in 
Fig.~\ref{soft}{\color{blue}}. 

\begin{figure}[!htb]
\centering
{\includegraphics[scale=0.33]{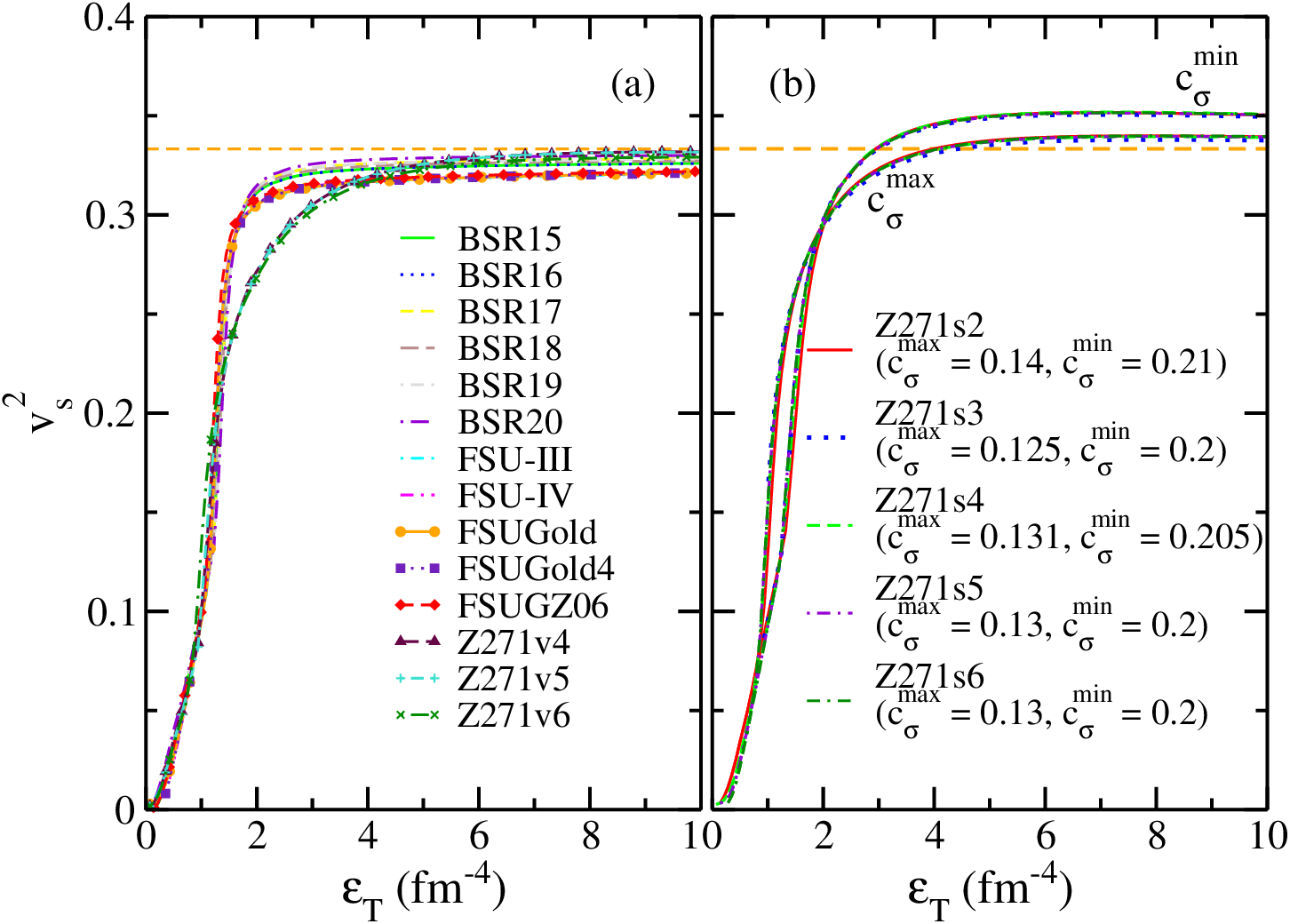}}
\caption{(Color online) Squared sound velocity for nuclear matter in beta 
equilibrium for (a) BSR, FSU, Z271v, and (b) Z271s parametrizations in the 
\mbox{$\sigma$-cut} scheme. Dashed line represents the limit $v^2_s = \frac{1}{3}$. 
The values of $c_\sigma$ for models in panel (a) are the same used in Fig.~\ref{soft}. In 
panel (b), the values of $c_\sigma^{min}$ and $c_\sigma^{max}$ are those that produce 
maximum neutron star mass of $1.93M_\odot$ and $2.05M_\odot$, respectively.}
\label{sound-soft}
\end{figure}

As one can see, the maximum neutron star mass constraint is now satisfied. Moreover, if 
we further investigate these modified parametrizations under the sound velocity bound 
proposed in Ref.~\cite{paulo}, we see that within the \mbox{$\sigma$-cut} scheme, they are
now consistent with such a constraint, as showed in 
Fig.~\ref{sound-soft}{\color{blue}(a)}.

However, we see in Fig.~\ref{sound-soft}{\color{blue}(b)} that for the Z271s 
parametrizations the sound velocity bound is not verified even within the 
\mbox{$\sigma$-cut} scheme. This is true for any $c_\sigma$ inside a range of 
$c_\sigma^{min}\leqslant c_\sigma \leqslant c_\sigma^{max}$, where $c_\sigma^{min}$ 
($c_\sigma^{max}$) is the value that produces a maximum neutron star mass equal to 
$1.93M_\odot$ ($2.05M_\odot$) for the Z271s models. In order to try to
understand the origin of such a result, we point out to the reader that models analyzed in 
Fig.~\ref{sound-soft} are described by the Lagrangian density given by 
$\mathcal{L}=\mathcal{L}_{Walecka}- \frac{A}{3}\sigma^3 - \frac{B}{4}\sigma^4+ 
\frac{C}{4}(g_\omega^2\omega_\mu\omega^\mu)^2+\mathcal{L}_{\sigma\omega\rho}$, where 
$\mathcal{L}=\mathcal{L}_{Walecka}$ is the very known Lagrangian density of the linear 
Walecka model (see Ref.~\cite{PRC90-055203}), and $\mathcal{L}_{\sigma\omega\rho} = 
g_\sigma g_\omega^2\sigma\omega_\mu\omega^\mu \left(\alpha_1+\frac{1}{2}{\alpha_1}
'g_\sigma\sigma\right)+ g_\sigma g_\rho^2\sigma\vec{\rho}_\mu\vec{\rho}^\mu 
\left(\alpha_2+\frac{1}{2}{\alpha_2} 'g_\sigma\sigma\right) + 
\frac{1}{2}{\alpha_3}'g_\omega^2 
g_\rho^2\omega_\mu\omega^\mu\vec{\rho}_\mu\vec{\rho}^\mu$. For the Z271s models not 
satisfying the sound velocity bound, there is no other isoscalar-vector interaction 
term besides the one regulated by the coupling constant $C$. In these models, 
$\alpha_1={\alpha_1}'={\alpha_3}'=0$ (also $\alpha_2=0$). All the other models consistent 
with the limit of $v_s^2$ in Fig.~\ref{sound-soft}{\color{blue}(a)}, present at least the 
term whose strength is controlled by the constant ${\alpha_3} '$, besides that one 
presenting $C$. Actually, this is the case for the Z271v models. They have  
$\alpha_1={\alpha_1}'=\alpha_2={\alpha_2}'=0$, but only $C$ and ${\alpha_3}'$ not equal to 
zero. One can see that the isoscalar-vector interaction seems to play an important role 
in the phenomenology of RMF models, concerning the simultaneous consistency of the sound 
velocity bound of $v_s^2=\frac{1}{3}$, verified in Ref.~\cite{paulo}, and the maximum 
neutron star mass constraint given by \mbox{$1.93\leqslant M/M_\odot\leqslant 2.05$}.

\subsection{Hyperonic matter}

We now consider the case when hyperons are also included in the EoS. At this point, we 
remind the reader that the inclusion of hyperons in hadronic RMF models is very 
important in the context of compact stars, since simple energetic 
considerations~\cite{Glen,Glen2} suggest they should be present at the high density 
regime that such objects naturally exhibit.  For any hadronic system, as the baryon 
density increases, the Fermi level rises sufficiently to start allowing for the emergence 
of the baryonic components presenting strangeness, i. e., hyperons such as $\Lambda$, 
$\Sigma^+$, $\Sigma^0$, $\Sigma^-$, $\Xi^0$, and $\Xi^-$. Due to the inclusion of these 
hyperons, the equation of state becomes softer with a consequent reduction of the maximum 
neutron star mass~\cite{Glen,Glen2}. However, the current status of this observational 
quantity points out to the opposite direction, i. e., recent findings of the massive stars 
\mbox{PSR J1614-2230} and \mbox{PSR J0348+0432} show that old predictions for the maximum 
neutron star mass around $1.44M_\odot$, are now replaced by the range of 
\mbox{$1.93\leqslant M/M_\odot\leqslant2.05$}, as we considered in our work. This fact 
leads to the problem that the EoS of the models needs to be stiffer in order to predict 
higher masses and this is not the case if hyperons are included. This contradiction is 
known in the literature as the {\it hyperon-puzzle}, and nowadays, many efforts have been 
directed to this subject in order to circumvent it as seen in Refs.~\cite{hyperons, 
james2015}.

Regarding the meson-hyperon couplings, they are responsible for large differences in 
similar calculations for stellar masses with hyperons~\cite{Glen, Weiss, luiz2014}. It 
is not the purpose of the present work to discuss this point in detail since a 
comprehensive analysis is not simple~\cite{james2015}, but we make two choices so that the 
results can be better understood. We define the ratio between the meson-hyperon and the 
meson-nucleon couplings as $\chi_\sigma = \frac{g_{Y\sigma}}{g_{N\sigma}}$, $\chi_\omega 
=\frac{g_{Y\omega}}{g_{N\omega}}$, $\chi_\rho =\frac{g_{Y\rho}}{g_{N\rho}}$, $\chi_\delta 
= \chi_\sigma$, where $Y$ represents any of the six lowest mass hyperons and $g_{Ni}$ 
represents the coupling of the nucleon with any of the four fields and it is density 
dependent in the cases we study next. The dependence with the density varies according to 
model used and they can be found in the original references given in Table~\ref{tab1}.

The first scenario we examine is the one called {\it universal coupling}~\cite{PRD9-1613}, 
given by $\chi_\sigma=\chi_\omega=\chi_\rho=\chi_\delta=1$ and we next call it SET~$1$. 
For the second scenario, considered more realistic because it is based on hypernuclei 
experimental values~\cite{Glen}, we use $\chi_\sigma=\chi_\delta=0.7$, 
$\chi_\omega=\chi_\rho=0.783$ and we call it SET~$2$. The results are displayed in 
Table~\ref{tab2}. 
\begin{table}[!htb]
\scriptsize
\caption{Neutron star main properties with hyperonic matter}
\centering
\begin{tabular}{l|c|c|c|c|c}
\hline\hline
couplings & Model & $M_{\rm max}/M$$_\odot$ & $R$~(km) & $R_{1.44M_\odot}$~(km) & 
$\varepsilon_c$(fm$^{-4}$) \\
\hline
SET 1 & TW99 & $1.895$ & ~~$9.631$ & $11.015$ & $8.671$\\
\hline
SET 2 & TW99 & $1.700$ & $10.166$ & $11.432$ & $7.774$ \\
\hline
\hline
SET 1 & DDH$\delta$ & $2.302$ & $11.307$ & $12.523$ & $5.781$ \\ 
\hline
SET 2 & DDH$\delta$ & $2.364$ & $12.491$ & $13.532$ & $4.485$ \\
\hline
\hline
SET 1 &DD-ME$\delta$ & $2.173$ & $10.420$ & $11.283$ & $7.005$ \\
\hline 
SET 2 & DD-ME$\delta$ & $2.254$ & $11.473$ & $12.352$ & $5.681$ \\
\hline\hline
\end{tabular}
\label{tab2}
\end{table}
One can see that the maximum masses obtained with SET~$2$ are higher if the $\delta$ 
meson is present in the model and lower otherwise. The inclusion of hyperons decrease the 
maximum mass as compared with the results with nucleons only by $9\%$ to $19\%$, depending
on the choice for the meson-hyperon constants. It is important to stress that many other 
choices would be possible, but in some cases the baryon effective masses decrease very 
rapidly and the EoS stops converging before the maximum stellar masses are attained. The
convergence of the codes should always be carefully checked. In the cases under 
examination, the radii of the canonical $1.44M_\odot$ star remain within the expected 
range. 

\section{Final remarks}

In the present paper we have revisited the  RMF models that were shown to satisfy 
several nuclear matter constraints in~\cite{PRC90-055203} and confronted them with 
astrophysical constraints. From the 35 analyzed models with nucleonic matter 
included, only the BKA20, BKA22, BKA24, BSR8, BSR9, BSR10, BSR11, BSR12, FSUGZ03, G2*, 
IU-FSU, and DD-F describe neutron stars with maximum mass in the range of 
\mbox{$1.93\leqslant M/M_\odot\leqslant 2.05$}~\cite{nature467-2010,science340-2013}. 
Only three models can sustain maximum masses larger than $2.05M_\odot$ when nucleonic 
matter is considered (TW99, DDH$\delta$, \mbox{DD-ME$\delta$}) and only two still reach 
this value once hyperons are included (DDH$\delta$, \mbox{DD-ME$\delta$}). These two 
models have density dependent couplings and $\delta$ mesons in their Lagrangian density. 

A possible alternative to make the EoS stiffer without altering its properties below 
nuclear matter saturation density was proposed recently in~\cite{voskre2015}.  Such an 
alternative could be used in order to {\it save} the RMF models presenting a maximum mass 
below $1.93M_\odot$, inside the constraint of \mbox{$1.93\leqslant M/M_\odot\leqslant 
2.05$}. The idea is to force the $\sigma$ self-interaction potential to rise sharply 
around densities just a bit larger than nuclear saturation density, resulting in the 
increase of the maximum possible stellar mass. As an example, the authors 
of~\cite{voskre2015} use the FSUGold model~\cite{PRL95-122501}, one of the models approved 
in our extensive analysis with stellar macroscopic properties shown in Table~\ref{tab1}. 
The maximum mass was increased from $1.72M_\odot$ to $2.01M_\odot$. Another more standard 
possibility is to consider also strange meson fields in the Lagrangian 
density~\cite{Weiss,luiz2014} and choose the hyperon-meson couplings such that the 
appearance of strange hyperons are pushed toward high densities. It is also important to 
remind the reader that a particular procedure of modifying RMF models in order to make 
them appropriate to describe neutron stars, without changing their bulk parameters at the 
saturation density, was  proposed in Ref.~\cite{serot} a long time ago. This procedure is 
basically supported by the inclusion of quartic self-interactions of the vector-isoscalar 
and vector-isovector fields related to $\omega$ and $\rho$ mesons, respectively. The 
strength of such interactions, not fixed at the saturation density, but free to run, can 
stiffen or soften the EoS, producing neutron stars whose maximum masses differ by more 
than one solar mass~\cite{serot}.

As far as the sound velocity in dense matter is concerned, our study corroborates the 
conclusions reached in~\cite{paulo}, namely, models that satisfy correct low density 
properties  and generate maximum star masses around $2M_\odot$, produce squared sound 
velocities larger than $1/3$ still at densities present in the stellar core. However, we 
verified that when the \mbox{$\sigma$-cut} scheme proposed in Ref.~\cite{voskre2015} is 
applied in those RMF models presenting maximum neutron star mass below the range of 
\mbox{$1.93\leqslant M/M_\odot\leqslant 2.05$}, it makes consistent such a constraint with 
the sound velocity bound of Ref.~\cite{paulo}. We also found that the isoscalar-vector 
interaction is an important ingredient in order to validate this consistency.

We have also analysed the possibility of fast cooling induced by the direct Urca process 
and have checked that 32 out of 34 models could give rise to it. 

As a last remark, we point out to the reader two recent RMF parametrizations proposed 
in~\cite{agrawal}, namely, BSP and IUFSU*. They were shown to predict maximum stellar 
masses inside the range of \mbox{$1.93\leqslant M/M_\odot\leqslant 2.05$} and also to 
allow the DU process. For the sake of completeness, we have submitted such models to 
the constraints of~\cite{PRC90-055203} and verified that they were approved together with 
the parameterizations analyzed in this work. We have obtained their central energy 
densities and the proton fraction related to the threshold of the DU process. The values 
are $\varepsilon_c=6.688$~fm$^{-4}$ and $Y_{\rm DU}=0.139$ for BSP, and 
$\varepsilon_c=6.141$~fm$^{-4}$ and $Y_{\rm DU}=0.138$ for IUFSU*. We also verified that 
for both parametrizations, the bound $v_s^2 < 1/3$ is violated at central energy densities 
smaller than $\varepsilon_c$, corroborating once more the findings of~\cite{paulo}.

\section*{Acknowledgments} 
This work was partially supported by CNPq (grants 300602/2009-0) and FAPESC (Brazil) 
under project 2716/2012, TR 2012000344. O.L. also acknowledges CNPq, and the support of 
Grant No. 2013/26258-4 from S\~ao Paulo Research Foundation (FAPESP). M.D. acknowledges 
support from Funda\c{c}\~ao de Amparo \`a Pesquisa do Estado do Rio de Janeiro (FAPERJ), 
Grant No. 111.659/2014.

\end{document}